\documentclass[twoside,twocolumn,9pt]{article}
\usepackage{extsizes}
\usepackage[super,sort&compress,comma]{natbib} 
\usepackage[version=3]{mhchem}
\usepackage[left=1.5cm, right=1.5cm, top=1.785cm, bottom=2.0cm]{geometry}
\usepackage{balance}
\usepackage{mathptmx}
\usepackage{sectsty}
\usepackage{graphicx} 
\usepackage{lastpage}
\usepackage[normalem]{ulem}
\usepackage[format=plain,justification=justified,singlelinecheck=false,font={stretch=1.125,small,sf},labelfont=bf,labelsep=space]{caption}
\usepackage{float}
\usepackage{fancyhdr}
\usepackage{fnpos}
\usepackage[english]{babel}
\addto{\captionsenglish}{%
  
}
\usepackage{array}
\usepackage{droidsans}
\usepackage{charter}
\usepackage[T1]{fontenc}
\usepackage[usenames,dvipsnames]{xcolor}
\usepackage{setspace}
\usepackage[compact]{titlesec}
\usepackage{hyperref}

\usepackage{amsmath}
\usepackage{dsfont}
\usepackage{amsfonts}

\newcommand{\VEC}[1]{\mathbf{#1}}
\newcommand{\HAT}[1]{\hat{\mathbf{#1}}}

\DeclareMathOperator{\tr}{tr}

\definecolor{darkpurple}{RGB}{150,0,200}

\usepackage{epstopdf}

\definecolor{cream}{RGB}{222,217,201}

\begin{document}

\pagestyle{fancy}
\thispagestyle{plain}
\fancypagestyle{plain}{
\renewcommand{\headrulewidth}{0pt}
}

\makeFNbottom
\makeatletter
\renewcommand\LARGE{\@setfontsize\LARGE{15pt}{17}}
\renewcommand\Large{\@setfontsize\Large{12pt}{14}}
\renewcommand\large{\@setfontsize\large{10pt}{12}}
\renewcommand\footnotesize{\@setfontsize\footnotesize{7pt}{10}}
\makeatother

\renewcommand{\thefootnote}{\fnsymbol{footnote}}
\renewcommand\footnoterule{\vspace*{1pt}%
\color{cream}\hrule width 3.5in height 0.4pt \color{black}\vspace*{5pt}} 
\setcounter{secnumdepth}{5}

\makeatletter 
\renewcommand\@biblabel[1]{#1}            
\renewcommand\@makefntext[1]%
{\noindent\makebox[0pt][r]{\@thefnmark\,}#1}
\makeatother 
\renewcommand{\figurename}{\small{Fig.}~}
\sectionfont{\sffamily\Large}
\subsectionfont{\normalsize}
\subsubsectionfont{\bf}
\setstretch{1.125} 
\setlength{\skip\footins}{0.8cm}
\setlength{\footnotesep}{0.25cm}
\setlength{\jot}{10pt}
\titlespacing*{\section}{0pt}{4pt}{4pt}
\titlespacing*{\subsection}{0pt}{15pt}{1pt}

\fancyfoot{}
\fancyfoot[RO]{\footnotesize{\sffamily{1--\pageref{LastPage} ~\textbar  \hspace{2pt}\thepage}}}
\fancyfoot[LE]{\footnotesize{\sffamily{\thepage~\textbar\hspace{.05 cm} 1--\pageref{LastPage}}}}
\fancyhead{}
\renewcommand{\headrulewidth}{0pt} 
\renewcommand{\footrulewidth}{0pt}
\setlength{\arrayrulewidth}{1pt}
\setlength{\columnsep}{6.5mm}
\setlength\bibsep{1pt}

\makeatletter 
\newlength{\figrulesep} 
\setlength{\figrulesep}{0.5\textfloatsep} 

\newcommand{\topfigrule}{\vspace*{-1pt}%
\noindent{\color{cream}\rule[-\figrulesep]{\columnwidth}{1.5pt}} }

\newcommand{\botfigrule}{\vspace*{-2pt}%
\noindent{\color{cream}\rule[\figrulesep]{\columnwidth}{1.5pt}} }

\newcommand{\dblfigrule}{\vspace*{-1pt}%
\noindent{\color{cream}\rule[-\figrulesep]{\textwidth}{1.5pt}} }

\makeatother

\twocolumn[
  \begin{@twocolumnfalse}
  \sffamily
   \noindent\LARGE{\textbf{
Preprint}}
{\includegraphics[height=0pt]{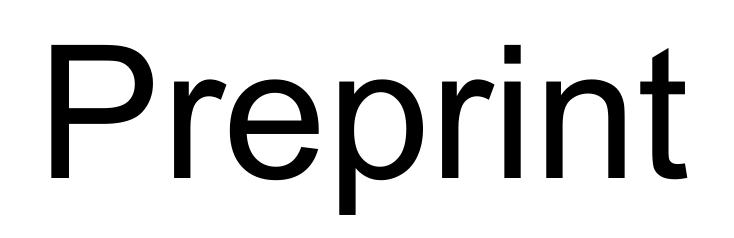}\hfill\raisebox{0pt}[0pt][0pt]{\includegraphics[height=55pt]{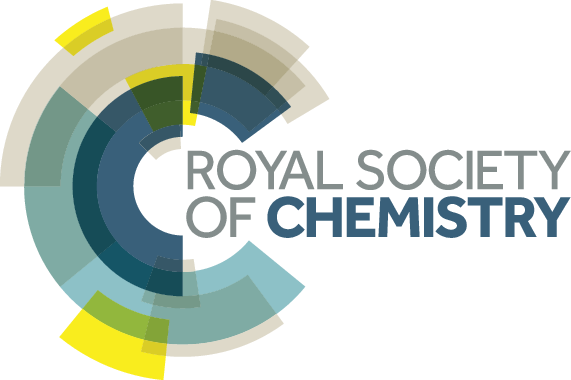}}\\[1ex]
\includegraphics[width=18.5cm]{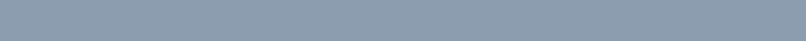}}\par
\vspace{1em}

\sffamily
\begin{tabular}{m{2. cm} p{13.5cm} }
{\includegraphics[height=0pt]{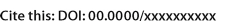}} & \noindent\LARGE{\textbf{
Continuum description of confluent tissues with spatial heterogeneous activity}}\\
\vspace{0.6cm} & \vspace{0.6cm} \\

 & \noindent\large{Fernanda P\'erez-Verdugo $^{\ast}$\textit{$^{a}$} and Rodrigo Soto\textit{$^{b}$}} \\

{\includegraphics[height=0pt]{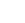}} & \noindent\normalsize{ A continuum description is built to characterize the stationary and transient deformations of confluent tissues subject to heterogeneous activities. By defining a coarse-grained texture matrix field to represent the shape and size of cells, we derive the coarse-grained stress tensor for the vertex model. Activity in the tissue takes the form of inhomogeneous apical contractions, which can be modeled as reductions of the vertex model reference areas or perimeters representing activity in the medial and perimeter regions of the cells, respectively. For medial activity, the extra stress is just an isotropic pressure, while for perimeter activity, it also has a deviatoric component, which is aligned with the texture matrix. The predictions of the continuum description are compared with the average spatiotemporal deformations obtained in simulations of the vertex model subject to localized apical contractions, showing an excellent agreement, even if the active patch is as small as one cell. The fluctuations around the average are more prominent when the activity is in the medial region due to the lack of negative active shape feedback, which, coupled with the confluent property, increases cellular shape and size variations.

} \\

\end{tabular}

 \end{@twocolumnfalse} \vspace{0.6cm}

  ]

\renewcommand*\rmdefault{bch}\normalfont\upshape
\rmfamily
\section*{}
\vspace{-1cm}

\footnotetext{\textit{$^{a}$~Department of Physics, Carnegie Mellon University,  Pittsburgh, PA 15213, USA.\\ E-mail: fverdugo@andrew.cmu.edu}}
\footnotetext{\textit{$^{b}$~Departamento de F\'{\i}sica, FCFM, Universidad de Chile, Santiago, Chile.}}





\section{Introduction}
Heterogeneity is detected across all scales in biological systems, from cellular components to organs. Variations in morphology \cite{burke1996phenotypic,aird2012endothelial}, expression levels \cite{tonotsuka2006heterogeneous}, and function \cite{aird2012endothelial} are particularly observed across cells within a tissue. 
For a long time, the average response has been researched as a representation of these systems.
Yet recent studies have emphasized the biological significance of heterogeneity in various facets, such as cancer \cite{lin2019biological, alizadeh2015toward}, epithelial wound healing \cite{vishwakarma2020dynamic}, and cell competition \cite{li2019mechanical}, highlighting the need for comprehensive knowledge and measurements of (population, temporal, and spatial) variations in biological systems \cite{gough2017biologically}.

Non-genetic heterogeneity, also known as phenotypic heterogeneity, can be induced by extrinsic and intrinsic variables in epithelial tissues \cite{gough2017biologically}, and has been associated with changes in cell-level mechanical properties \cite{vishwakarma2020dynamic,tsuboi2017inference}. Variations in cellular adhesion and contractility specifically govern cell sorting or mixing \cite{tsuboi2017inference}. In the context of cancer, heterogeneity is of special importance since intratumor areas exhibit high morphological variations linked with adaptation and resistance to therapy, hence promoting disease progression \cite{lin2019biological, alizadeh2015toward}.

Some recent cell-level computational and continuum models account for tissue heterogeneities \cite{tsuboi2017inference,li2019mechanical,murphy2020mechanical}. Specifically, in Ref.~\cite{li2019mechanical}, the authors analyzed population heterogeneity in relation to mechanical heterogeneity using a cell-level model, showing that variations in the cellular shape index (defined as the ratio of the preferred perimeter to the square root of the preferred area) increase tissue-level rigidity. Similarly, in Ref.~\cite{perez2022geometrical}, we considered mechanical heterogeneities, localized in space and time,  as  active processes associated with internal changes at specific cellular regions, medial and periphery, resulting in variations in the cellular shape index. Thus, we were able to make quantitative predictions on the stress localization based on the observed experimental cellular morphologies. The intermediate spatial scale of these inhomogeneous activities, which are directly linked to large-scale mechanical inhomogeneities, has not been studied using numerical simulations or continuous models.

In this paper, we present a continuum description for confluent tissues in the absence of cellular rearrangements, exhibiting spatially inhomogeneous cellular activity that is directly related to mechanical heterogeneities. The continuum theory is founded on a coarse-grained expression for the tissue stresses, derived from the vertex model, as previously reported in Ref.~\cite{ishihara2017cells}, with a direct mapping between the cell-level and continuum parameters. We compare the outcomes of the continuum description with numerical simulations employing the vertex model under various scenarios where activity causes inhomogeneous cellular contractions exhibiting fluidity or solidity geometrical features. Even for active regions as small as one cell, the agreement between the two descriptions is excellent. Our validated continuous model offers a substantial analytical and numerical advantage. It allows, for example, the direct inclusion of spatial symmetries that describe the tissue dynamics, such as during the formation of the ventral furrow formation in \textit{Drosophila} \cite{spahn2013vertex,heer2017actomyosin}, and the neural tube in \textit{Xenopus} \cite{inoue2016mechanical}.
We finally discuss the effect of intrinsic disorder in the tissue, included as a population heterogeneity in numerical simulations (as in Ref.~\cite{li2019mechanical}), on the fluctuations of the microscopic model around the predictions of the continuum description. 


\section{Continuum description of the vertex model} \label{sec.continuum}

The two-dimensional vertex model  \cite{nagai1988vertex,nagai2001dynamic} has been demonstrated to be both qualitatively and quantitatively adequate for describing the dynamics of morphogenic processes in confluent tissues \cite{farhadifar2007influence,fletcher2014vertex,kursawe2015capabilities}. In this model, each cell $c$ is represented by a polygon with area $A_c$ (target area $A_{0c}$) and perimeter $P_c$ (target perimeter $P_{0c}$). The position of the vertex $\VEC r_i$ follows a variational dynamics, $\mu {{\rm d} \VEC r_i} / {\rm d}t = -{\partial{E}}/{\partial{{\VEC r_i}}}$, where $\mu$ is the friction coefficient, and $E$ is an energy function given by
\begin{align}
E = \frac{K_A}{2}\sum_c\left( A_c - A_{0c}\right)^2 + \frac{K_P}{2}\sum_c\left( P_c -
P_{0c}\right)^2+J\sum_{\langle i,j \rangle} l_{ij}.
\label{eq.Evertex}
\end{align}
The first and second sums run over all the cells and penalize deviations of areas and perimeters, with elastic modulus $K_A$ and $K_P$, respectively. The third sum is taken over the adjacent vertices $i$ and $j$, joined by a cell junction of length $l_{ij}$ under a constant line tension $J$. Although it is possible to absorb the constant line tension $J$ into the perimeter penalization term, we keep all three terms as in Ref.~\cite{sato2021novel}, and directly denominate medial (perimeter) activity to spatial variations of $A_{0c}$ ($P_{0c}$).

A useful way to characterize the shape of a cell $c$, made of $n_c$ vertices with positions $\VEC{r}_i$, is given by the symmetric texture tensor 
 \begin{align}
\mathbb{M}_c =\frac{2}{n_c}\sum_{i\in c}\left( \VEC{r}_i - \VEC{r}_c\right)\otimes \left( \VEC{r}_i - \VEC{r}_c\right),
 \label{eq.M}
\end{align}
where $\VEC{r}_c$ is the centroid of the cell $c$ \cite{nestor2018relating}. For a regular hexagon, $\mathbb{M}_c$ is equal to $R_{\text{out}}^2\mathds{1}$, where $R_{\text{out}}$ corresponds to the circumscribed circle and $\mathds{1}$ is the identity matrix. For a rhombus of diagonals $2a$ and $2b$, with $a>b$ (low-resolution version of an ellipse of semi-axis $a$ and $b$), the eigenvalues of $\mathbb{M}_c$ are $a^2$ and $b^2$, and the eigenvectors give the orientations of each diagonal. In general, the area and perimeter of the cell $c$ can be written as $A_c=c_1 \sqrt{\det \mathbb{M}_c}$ and $P_c=c_2\sqrt{\tr {\mathbb{M}_c}}$, where $c_1$ and $c_2$  depend on the type of polygon. In particular, for the previous hexagon and rhombus, $c_1={\lbrace 3\sqrt{3}/2, 2\rbrace}$ and $c_2={\lbrace  3\sqrt{2}, 4\rbrace}$, respectively. For a hexagon of initial side $R$, under a pure shear deformation of amplitude $\epsilon$, the previous expression for the perimeter is corrected by  $-3R\epsilon^2/2+\mathcal{O}(\epsilon^4)$, and  the area by $-3\sqrt{3}R^2 \epsilon^2 +\mathcal{O}(\epsilon^4)$. For a square (a particular case of rhombus) of initial side $R$, under a horizontal expansion of amplitude $\epsilon$ (rectangle of sides $R$ and $R(1+\epsilon)$), the correction for the perimeter is $-R\epsilon^2/2+\mathcal{O}(\epsilon^3)$, while there is no correction for the area. Other deformations produce similar results. As a consequence, the previous expressions for $A_c$ and $P_c$ give the correct values to the dominant order, allowing us to use them in a continuous description.

 We consider a coarse-graining over the discrete perspective of the tissue in order to obtain a smooth, symmetric tensor field $\mathbb{M}(\VEC{r})$, which now reflects shape at a tissue scale. Therefore, as shown in Refs.~\cite{ishihara2017cells}, under a displacement field $\VEC u$, the tensor field $\mathbb{M}$ will be modified (up to $O(\vert \nabla \VEC u\vert )$) as
 \begin{align}\mathbb{M}_{ij}^{\text{new}} = \mathbb{M}_{ij}-u_k (\partial_k \mathbb{M}_{ij})+(\partial_k u_i)\mathbb{M}_{kj}+\mathbb{M}_{ik}(\partial_k u_j),
 \label{eq.confluence} 
 \end{align}
ensuring the confluence of the tissue. Additionally, a free energy density can be written in terms of  $\mathbb{M}$, as

\begin{align}
f\left(\mathbb{M}\right)=&\frac{1}{c_1 \sqrt{\det{\mathbb{M}}}} \Bigg\{\frac{K_A}{2} \left(c_1 \sqrt{\det{\mathbb{M}}}-A_0\right)^2+
\nonumber \\ & 
\frac{K_P}{2} \left[c_2\sqrt{\tr{\mathbb{M}}}-\left(P_{0}-\frac{J}{2K_P}\right)\right]^2\Bigg\}
 \label{eq.f.maintext}
\end{align}
such that the total elastic energy \eqref{eq.Evertex} is obtained as $E=\int f\left(\mathbb{M}\left(\VEC r\right)\right){\rm d}^2 r$, with the areas an perimeters obtained from $\mathbb{M}$,  as in Ref.~\cite{ishihara2017cells} (see App.~\ref{app.stress} for details). See Refs. \cite{grossman2022instabilities} and \cite{hernandez2022anomalous} for other similar $2\times2$ symmetric tensors used to describe cell shapes in continuum models of tissues.

Since the tensor field $\mathbb{M}\left(\VEC r\right)$ is symmetric, it can be represented as $\mathbb{M} = M e^{c\Theta(\theta)}$, where $M$, $c$ and $\theta$ are scalar fields. $M$ represents the coarse-grained cellular area (over $c_1$); $c$, the coarse-grained cell shape anisotropy; and $\theta$, the coarse-grained cell orientation, in term of which the trace-less tensor $\Theta$ is given by 
 \begin{align}
\Theta=\begin{pmatrix}
\cos2\theta& \sin 2\theta \\
\sin2\theta & -\cos2\theta
\end{pmatrix}.
 \label{eq.theta}
\end{align}
In this representation, $c=\ln(s_1/s_2)$, where $s_1$ and $s_2$ are the semi-axis of the approximated ellipse-shape, with $s_1$ oriented along  $\theta$. Then, $c>0$ ($c<0$) represents cellular elongation (contraction) in the direction defined by $\theta$.

With the previous representation of $\mathbb{M}$, its determinant and trace are given by $\det{\mathbb{M}}=M^2$ and $\tr{\mathbb{M}} = 2M\cosh\left(c\right)$. Following the procedure shown in Ref.~\cite{ishihara2017cells} using  Eq.~\ref{eq.confluence} , it is possible to derive the associated elastic stress tensor given by a combination of an isotropic pressure and a deviatoric tensor, related to the tension at the cellular junctions: $\sigma_e = -p_e\mathds{1} + \sigma_{e,\text{dev}}$ (App.~\ref{app.stress}), with
 \begin{align}
&p_e = -K_A\left(c_1 M -A_0\right)-  K_P\frac{c_2}{c_1}\left[c_2 -\frac{P_0 - J/(2K_P)}{\sqrt{2M\cosh(c)}}\right]\cosh(c), \label{eq.p_e1}\\
&\sigma_{e,\text{dev}}=  K_P\frac{c_2}{c_1}\left[c_2 -\frac{P_0 - J/(2K_P)}{\sqrt{2M\cosh(c)}}\right]\sinh(c)\Theta.
 \label{eq.dev_e1}
\end{align}
We recall that the areas and perimeters are correctly computed in terms of $\mathbb{M}$ up to order $\epsilon$ in the deformations. Therefore, as Eqs.~\eqref{eq.p_e1} and \eqref{eq.dev_e1} linearly depend on the areas and perimeters, the stresses are correct also to linear order in the deformations. 

\section{Inhomogeneous activity} \label{sec.activitygeneral}

In Ref.~\cite{ishihara2017cells} they use a similar form of the previous stress tensor. In all their applications (externally induced axial stretch, deformation due to active internal forces, and generation of shear flow) they assume for simplicity that $\mathbb{M}$ is homogeneous, and the field $M$ is constant. Here, we consider the case of epithelial tissues performing inhomogeneous apical contractions, resulting in texture and stress tensors that are inhomogeneous.

In many circumstances, epithelial tissues exhibit dramatic alterations at the apical cellular surface. Particularly, the contraction of the apical face (reviewed in Ref.~\cite{martin2014apical}) is a fundamental mechanism of tissue remodeling. Apical contractions are seen during localized processes such as cell divisions or extrusions \cite{atieh2021pulsatile}, as well as during major coordinated cell movements such as ventral furrow formation in \textit{Drosophila} \cite{martin2020physical}, or the stage preceding epiboly in \textit{Austrolebias nigripinnis} \cite{reig2017extra}. 
In a previous article \cite{perez2022geometrical}, we addressed and computationally modeled the apical contraction of a single cell embedded in a tissue described with the vertex model, as an active process related to internal changes at specific cellular regions, medial and periphery. Similarly, other researchers have explored the mechanical responses in tubular epithelial systems \cite{okuda2017contractile} using the premise that all cells are actively contracting. They observed that cell-level geometrical changes generated by belt-like and mesh-like activities influence tube-level stiffness. On the other hand, Spahn \textit{et al.\ }\cite{spahn2013vertex} specifically analyzed the ventral furrow formation in \textit{Drosophila}, taking into account active energy terms proportional to cell areas and junction lengths to generate the apical contractions, and were able to predict the experimentally observed anisotropic constriction.

The microscopic description of the vertex model (at the level of individual cells and vertices) can be too detailed when describing large tissues, making it hard to make analytical progress. Furthermore, for tissues that present some spatial symmetries, it is not direct to theoretically impose those symmetries on the vertex displacements. It is, then, natural to apply the continuum description for the analysis of inhomogeneous apical contractions.

In the vertex model, medial and perimeter activities associated to apical contractions are represented by inhomogeneous modifications of the target quantities, as $A_{0} \rightarrow  A_{0} + \Delta_A\left(\VEC r\right)$ and $P_{0} \rightarrow P_{0}+ \Delta _P \left(\VEC r\right)$. Since the shape index here is given by $p_0=\left[P_0 - J/\left(2K_P\right)\right]/\sqrt{A_0}$, these inhomogeneous activities induce inhomogeneities in $p_0$, which is known to control fluidity in the classical vertex model \cite{bi2015density}, and its heterogeneity has been proven to enhance rigidity \cite{li2019mechanical}.
Using Eqs. \eqref{eq.p_e1} and \eqref{eq.dev_e1}, the stresses  associated with the activities are given by $\sigma_{A_0} = -p_{A_0}\mathds{1}$ and $\sigma_{P_0}  = -p_{P_0}\mathds{1} + \sigma_{P_0,\text{dev}}$, with

\begin{align}
& p_{A_0} = \Delta_A K_A ,\label{eq.p_A0}  \\
&p_{P_0} = \frac{\Delta_P}{\sqrt{2M}} K_P \frac{c_2}{c_1} \sqrt{\cosh\left(c\right)}  ,\label{eq.p_P0}  \\
&\sigma_{P_0,\text{dev}} = - \frac{\Delta_P}{\sqrt{2M}}  K_P \frac{c_2}{c_1} \frac{\sinh\left(c\right)}{ \sqrt{\cosh\left(c\right)}}\Theta \label{eq.dev_P0}.
\end{align}

While medial activity generates a pressure field $p_{A_0}$ (explicitly) independent of cell size ($M$) and shape ($c$), perimeter activity generates a pressure $p_{P_0}$ and a deviatoric stress $\sigma_{P_0,\text{dev}}$, both of them with inverse relation with size ($\sim 1/\sqrt{M}$), and dependent on shape. Both active pressures take the sign of the active term $ \Delta_{A,P}$. Instead, the sign of the active deviatoric stress $\sigma_{P_0,\text{dev}}$ depends on both $\Delta_{P}$ and $c$. Then, negative $\Delta_{A,P}$ generates negative active pressures, causing cellular compression and local shape variations. Additionally, junctions will actively change their tension in the case of perimeter activity, depending on shape (elongation/contraction) and orientation. For example, for an initially isotropic tissue ($\theta=0$) under a negative activity that only depends on the horizontal position $\Delta_{P}=\Delta_{P}(x)<0$, elongated (contracted) junctions oriented in $\HAT x$ will actively increase (decrease) their tension, increasing shape isotropy, as shown at the single-cell activity level in Ref.~\cite{perez2022geometrical}. The latter implies that $\Delta_{P}<0$ locally solidifies the tissue, as expected in a system locally decreasing $p_0$. 

The tissue’s steady-state under these inhomogeneous activities is obtained imposing the force balance, $\nabla \cdot \sigma = 0$, i.e.,
\begin{align}
\nabla \left( p_e + p_{A_0} + p_{P_0}\right)= \nabla \cdot\left( \sigma_{e,\text{dev}} + \sigma_{P_0,\text{dev}}\right).
\label{eq.ss} 
\end{align}

To analyze the temporal evolution, we make an analogy to the equation of motion for an isotropic elastic medium \cite{landau1986theory}. We equate the internal  force coming from the total stress tensor $\nabla \cdot \sigma$, to the product of the velocity times the friction per unit area of the body (the vertex), i.e., $\tilde{\mu}=\mu/A_v$, where $A_v$ is the mean area occupied by a vertex. The resulting dynamical equation is then given by
\begin{align}\frac{\mu}{A_v}\frac{{\partial}\VEC u(\VEC r,t)  }{{\partial }t} = \nabla \cdot  \left[\sigma_e(\VEC r,t) + \sigma_{A_0}(\VEC r,t) + \sigma_{P_0}(\VEC r,t)\right].
\label{eq.temporal_stripe}
\end{align}
To summarize, the static and dynamic continuum descriptions are for the displacement field $\VEC u$, obeying the mechanical equations \eqref{eq.ss} and \eqref{eq.temporal_stripe} in terms of the stress tensor, which is computed from the texture field using equations \eqref{eq.p_e1} and \eqref{eq.dev_e1}. Finally, the texture field satisfies the confluence condition \eqref{eq.confluence} in terms of the displacement field, closing the system of equations. Activity provides additional stresses, Eqs. \eqref{eq.p_A0}-\eqref{eq.dev_P0}, driving the tissue to new configurations.

 
\section{Application: Simple active stripe} \label{sec.stripe}

To keep it simple but biophysically relevant, we apply the previous continuum description to the case of a stripe patch with a negative activity, which varies with a Gaussian profile inside the stripe, and compare results with numerical simulations using the vertex model (microscopic description).  This kind of activity causes cellular contractions as experimentally seen during the ventral furrow formation in \textit{Drosophila} \cite{martin2020physical,spahn2013vertex}. The case of a circular active patch, which produces apical geometries as seen in cyst formation \cite{bielmeier2016interface} and is more exigent to the theory, is analyzed App.~\ref{app.circular}.

Both for the vertex model and continuum descriptions, we set units by choosing $A_0=\langle A_{0c}\rangle=1$, $K_A=1$, and $\mu=1$, and define the dimensionless parameters $\hat{K}_P=K_P/(K_A A_0)$ and $\hat{J}=J/(K_A A_0^{3/2})$.

To obtain a mechanically stable tissue \cite{perez2020vertex} where no T1 events will be required in the numerical simulations, we chose the values $\hat{K}_P=\hat{J}=1$.

For the microscopic description,first we consider an initial two-dimensional ordered tissue made up of identical regular hexagonal cells, with uniform  $A_{0c}=1$  and $P_{0c}=3.72$, and then an isotropic disordered confluent tissue with non-uniform preferred quantities $A_{0c}$ and $P_{0c}$. To create the disordered tissue we built Voronoi cells where 3000 center points are generated by a Monte Carlo simulation of hard disks in a box of dimensions $L_x \times L_y$ (Table~\ref{tab.parameters_stripe}), with periodic boundary conditions. The diameter of the disks govern the degree of dispersion of the cells. We consider an area fraction $\phi=0.71$ \cite{perez2020vertex}, below the freezing transition, obtaining polygons with areas and perimeter that define $A_{0c}$ and $P_{0c}$ for each cell, with $\langle A_{0c}\rangle=1$ and $ \langle P_{0c}\rangle=3.9$ (see App.~\ref{app.sim.details} for more details). 
Particularly, we begin with a solid-tissue arrangement with mean shape index $ \langle p_{0c}\rangle =3.4$. The system is then allowed to relax using the parameters shown in Table~\ref{tab.parameters_stripe}, in the absence of activity. The relaxed state defines the initial configuration of our entire tissue, and has polygonal shapes from squares to nonagons, with mean area $\langle A_{c}\rangle=1$ and mean perimeter $ \langle P_{c}\rangle=3.77$ (App.~\ref{app.sim.details}). Note that the value of $\langle A_{c}\rangle$ is preserved during relaxation because the tissue is confluent, and hence the total area does not change.

\begin{figure}[t] 
\centering 
 \includegraphics[width=1\linewidth]{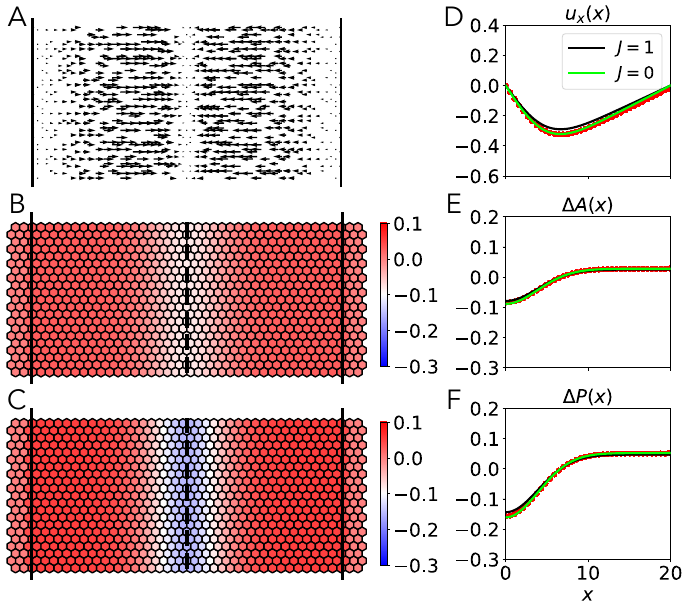} 
\caption{Contraction of a simple active stripe for regular hexagonal cells. Results of the simulated epithelial tissue made of initially regular hexagonal cells of area equals one, with $K_A=1$, $K_P=1$, $A_0=1$ , $P_0=3.72$, and $J=1$, under medial activity ($\lambda_A=0.5$, $\lambda_P=0$,  $R=3$, and $W_{\text{half-box}}=14$) at $t=100$, versus the initial position of the vertex or cell centers. (A) Vectorial map of the total vertex displacement in
a representative tissue section, in units of $\sqrt{A_0}/5$. (B)
Representative section of the tissue, showing area change. (C) Representative section of the tissue, showing perimeter change. (D) Scatter plot of the horizontal displacement of vertices (red dots). (E) Scatter plot of the area change (red dots). (F) Scatter plot of the perimeter change (red dots). 
In A-B-C the solid-black curves represent the distance $x=\pm W_{\text{half-box}}$. In D-E-F, the $x$ axis gives the initial position of the vertices or cell centers and the solid-black (lime) curves show the outcomes of the continuum description, with $J=1$ ($J=0$), $A_0=1$, $P_0=3.72$, $c_1=3\sqrt{3}/2$, and $c_2=3\sqrt{2}$.
}
\label{fig.hex-franja}
\end{figure}

\begin{figure}[t]
 \includegraphics[width=1\linewidth]{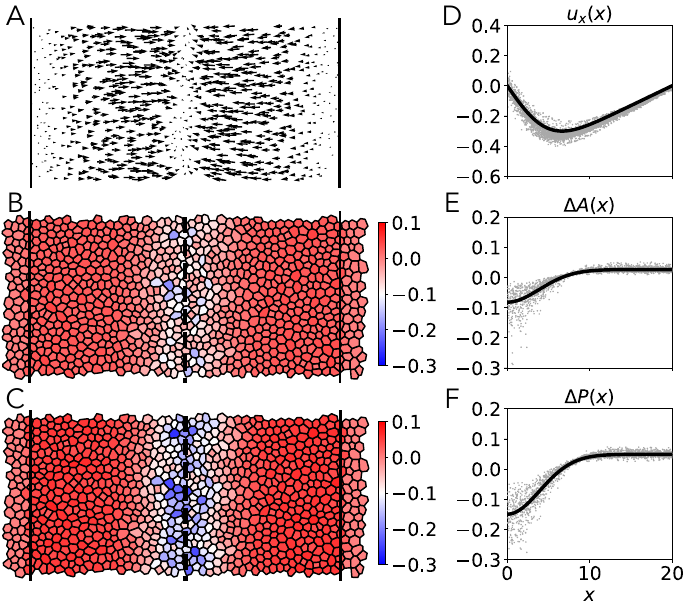} 
\caption{Contraction of a simple active stripe for a disordered tissue. Results of a simulated epithelial tissue with medial activity ($\lambda_A=0.5$, $\lambda_P=0$, $R=4$, $W_{\text{half-box}}=20$) at $t=100$.
(A) Vectorial map of the total vertex displacement in a representative tissue section, in units of $\left[\sqrt{A_0}/5\right]$. 
(B) Area change in a representative section of the tissue. 
(C) Perimeter change in a representative section of the tissue.
(D) Scatter plot of the horizontal displacement of each vertex. 
(E) Scatter plot of the area change. 
(F) Scatter plot of the perimeter change.
The dashed line in B-C shows the active axis ($x=0$). 
In A-B-C the solid-black curves represent the distance $x=\pm W_{\text{half-box}}$.
In D-E-F, the $x$ axis gives the initial position of the vertices or cell centers, and the solid-black curves show the outcomes of the continuum description.}
\label{fig.ejemplo_medial_R4}
\end{figure}

The active stripe is built as follows. 
We define an active axis along the $y$ direction, with a random $x$ position and we clamp all vertices that are further than $W_{\text{half-box}}=20$ away from this active axis (i.e.,  we impose no motion on these vertices). Then, we end up working with a smaller tissue made up of $\sim 2200$ cells that has periodic boundary conditions on the $y$-axis, and fixed boundary conditions on the $x$-axis. Medial and perimeter activities are included as $\Delta_A= -\lambda_A A_{0c}\Gamma(x) $ and $\Delta_P= -\lambda_P P_{0c}\Gamma(x) $, with $\Gamma(x)=\exp\left(-x^2 / 2R^2\right)$, and $R$ indicating the active region extension. These changes cause spatial-dependent cellular contractions, as seen in Fig.~\Ref{fig.ejemplo_medial_R4}. More specifically, the addition of medial (perimeter) activity locally raises (decreases) $p_{0c}$, to values that depend on the distance to the active axis.  
Finally, we let the system evolve by integrating the dynamical equation for ${\VEC r_i}$, using the Euler method with ${\rm d}t=0.05$, up to $t=100$, reaching a steady-state.


\subsection{Steady state} \label{sec.ss}

We analyze the final (steady) state of the simple active stripe simulations by quantifying the displacement of the vertices and the area and perimeter changes of the cells. Figs.~\ref{fig.hex-franja} (ordered tissue) and \ref{fig.ejemplo_medial_R4} (disordered tissue) show these three quantities for an active patch defined by $R=4$, under medial activity ($\lambda_A=0.5, \lambda_P=0$), with $x=0$ the position of the active axis (dashed-black lines), and $x$ the distance (in absolute value) from it. 
Figs.~\ref{fig.hex-franja}A (ordered tissue) and \ref{fig.ejemplo_medial_R4}A (disordered tissue)  show the vectorial map of $\mathbf{u}$, with horizontal and vertical components. For clarity, only the vector of one vertex per cell is shown. Small magnitudes of $\mathbf{u}$ (black arrows in units of $\left[\sqrt{A_0}/5\right]$) are obtained with the active parameters considered in this work, and then cells conserve their neighbors all the time. Initial geometrical disorder enhances local vertical components in the displacement, with positive and negative signs in different regions. However, in both tissues we obtain a horizontal net displacement when taking the average response of the vertices. Figs.~\ref{fig.hex-franja}B-C (ordered tissue) and \ref{fig.ejemplo_medial_R4}B-C (disordered tissue) show the color map of the obtained area changes $\Delta A$, and perimeter changes $\Delta P$, respectively. $\Delta A$ decreases near the active axis and increases away from it due to total area conservation. $\Delta P$ has the same qualitative behavior.

We proceed to analyze the steady state using the continuum description. For that, we consider an initial isotropic tissue with uniform $\mathbb{M}$, defined by $M(t=0) = m = A_0/c_1, c(t=0) = 0$ and $\theta(t=0)=0$. Since the activities depend on the horizontal position only, we use that spatial symmetry to impose that $\VEC u = u_x(x)\HAT x$, and $\theta(x)=0$ (contractions and elongations along the $x$-axis). Then, the confluence condition for the tissue [Eq.~\eqref{eq.confluence}] reads
\begin{align}
M\begin{pmatrix}
e^c & 0 \\
0 & e^{-c} 
\end{pmatrix} = m
\begin{pmatrix}
1+2\frac{{\rm d} u_x}{{\rm d}  x} & 0 \\
0 & 1
\end{pmatrix},
\end{align}
representing a functional relation between the scalar fields $M$ and $c$, and the displacement field $u_x(x)$, from where we obtain
\begin{align}
&M\left(x\right)=m\sqrt{1+2\frac{{\rm d} u_x}{{\rm d}  x}}, \label{eq.M0_u_x}\\
&\cosh\left[c\left(x\right)\right] = \frac{m}{M\left(x\right)} \left( 1+ \frac{{\rm d} u_x}{{\rm d}  x} \right)
= \frac{1+ \frac{{\rm d} u_x}{{\rm d}  x} }{\sqrt{1+2\frac{{\rm d} u_x}{{\rm d}  x}}} ,\label{eq.c1_u_x}\\
&\sinh\left[c\left(x\right)\right] = \frac{m}{M\left(x\right)} \frac{{\rm d} u_x}{{\rm d}  x}
= \frac{\frac{{\rm d} u_x}{{\rm d}  x}}{\sqrt{1+2\frac{{\rm d} u_x}{{\rm d}  x}}},\label{eq.c2_u_x}
 \end{align}
 and then $e^c =\cosh\left[c\left(x\right)\right] + \sinh\left[c\left(x\right)\right] =\sqrt{1+2\frac{{\rm d} u_x}{{\rm d}  x}}$.

The steady-state condition [Eq.~\eqref{eq.ss}] in this case is given by
  \begin{align}
  0  =& \frac{\partial \sigma_e^{xx}}{\partial x} + \lambda_A K_A A_0 \frac{\partial \Gamma(x)}{\partial x} +\lambda_P P_0 K_P \frac{c_2}{c_1} \frac{\partial}{\partial x}  \left[\frac{e^c \Gamma(x)}{\sqrt{2M \cosh(c)}} \right],
\label{eq.eq_x} 
\end{align}
where 
\begin{align}
\sigma_e^{xx}=& K_A\left(c_1M-A_0\right)+ K_P\frac{c_2}{c_1}\left(c_2 -\frac{P_0 - J/(2K_P)}{\sqrt{2M\cosh(c)}}\right)e^c.
\end{align}

Finally, we can write each term in Eq.~\eqref{eq.eq_x} in terms of $x$, $u_x\left(x\right)$,  $u_x'\left(x\right)$, and $u_x''\left(x\right)$, obtaining an ordinary differential equation which we numerically solve imposing $u_x(x=0)=0$, $u_x(x=W_{\text{half-box}})=0$. 

First, we compare the results from numerical simulation using the ordered tissue with the ones obtained from the continuum description when considering  $A_0=1$, $P_0=3.72$, $c_1=3\sqrt{3}/2$, and $c_2=3\sqrt{2}$ (hexagon values). As a result, we obtain $u_x(x)$ and use it to compute $\Delta A$ and $\Delta P$. Due to the symmetry of the regular hexagonal lattice, and the borders of the tissue being clamped, no net effect of $J$ over the tissue dynamics is expected. Indeed, we obtained that the results from the numerical simulations are independent of $J$, and hence saying $J=1$ is the same as $J=0$, for the ordered tissue. However, we find that the solution of our continuum description changes when varying $ J$ as seen in Fig.~\ref{fig.hex-franja}D-E-F (black and lime curves). Figs.~\ref{fig.circular-hexagons} and \ref{fig.S5} show the same result, under different lattices and symmetries. 
This is an effect of approximating areas and perimeters as being simply proportional to $\sqrt{\det \mathbb{M}}$ and $ \sqrt{\tr {\mathbb{M}}}$, respectively. 
Then, more complex expressions for the area an perimeter would be needed in order to recreate the tissue dynamical independence of $J$ in ordered tissues. Nevertheless, we will show that under our simple definitions, with shape begin defined by only two degrees of freedom (fields $M$ and $c$), we can very closely describe disordered tissues.

For the disordered tissue used in numerical simulations, since it is isotropic and has a majority of hexagons  (Fig.~\ref{fig.initial_state}), we keep the values $c_1=3\sqrt{3}/2$ and $c_2=3\sqrt{2}$, and consider initial areas and perimeters given by a hexagonal lattice ($A_{\text{hex}}=1$, $P_{\text{hex}}=3.72$). Figs.~\ref{fig.ejemplo_medial_R4}D-E-F show the results obtained by the continuum description (solid-black curves), together with $\Delta A$ and $\Delta P$ measured in the vertex model simulations using the disordered tissue. App.~\ref{app.circular} presents the case of an active circular region of Gaussian size $R=4$, for the ordered and disordered tissue.

\begin{figure}[t]
\centering
 \includegraphics[width=1\linewidth]{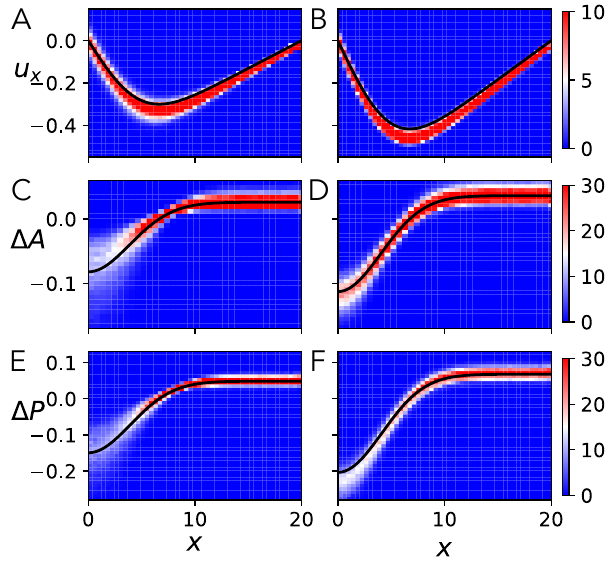} 
\caption{Statistical analysis of the tissue-level response on the simple active stripe. Probability density function by column, obtained from 100 simulations with randomly chosen active axis, at $t=100$, versus the initial position of the vertex or cell center. Left: simple active stripe with medial activity $\left(\lambda_A=0.5,\lambda_P=0, R=4\right)$. Right: simple active stripe with perimeter activity $\left(\lambda_A=0,\lambda_P=0.1, R=4\right)$. (A)-(B) Horizontal displacements of the vertices. (C)-(D) Cellular area change $\Delta A$. (E)-(F) Cellular perimeter change $\Delta P$. The solid-black curves show the outcomes of the continuum description.}
\label{fig.R4_desordenado}
\end{figure}

As seen in Fig.\ \ref{fig.ejemplo_medial_R4}, vertex model simulations using the disordered tissue show important fluctuations in the different observables. To quantify this, Fig.\  \ref{fig.R4_desordenado} shows the probability density function by column, of $u_x$, $\Delta A$, and $\Delta P$, for the cases of medial and perimeter activity, considering 100 different simulations (random active axis positions) for each case. Again, solid-black curves show the results obtained from solving the steady-state condition. In both descriptions, i) the maximum area and perimeter change ($\Delta A_{\text{max}},\Delta P_{\text{max}}$) is reached at the active axis ($x=0$); ii) the area change changes its sign due to the total area conservation, defining a region under contraction and another under expansion; iii) the displacement has a well-defined minimum (maximum displacement, $u_{\text{max}}$). Additionally, the geometrical response under medial activity presents more variability than under perimeter activity (Fig.~\ref{fig.R4_desordenado}).

\begin{figure}[t]
 \includegraphics[width=.85\linewidth]{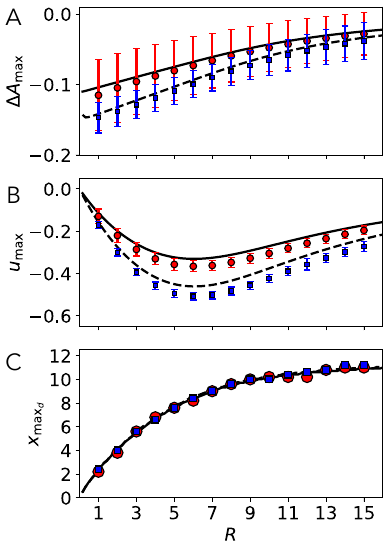} 
\caption{Comparison of micro and macro descriptions of the steady-state. (A) Maximum contraction for different active stripe sizes $R$. (B) Maximum horizontal displacement. (C) Position at which the horizontal displacement is maximum. Black curves show the outcomes of the continuum description, for medial $\left(\lambda_A=0.5,\lambda_P=0\right)$ (solid-line) and perimeter $\left(\lambda_A=0,\lambda_P=0.1\right)$ (dashed-line) activity. Red circles and blue squares represent the numerical results averaged over 100 simulations with different active axes for each integer $R$ from 1 to 15. Bars show the standard deviations of $\Delta A$ and $u$ at the position of maximum contraction and displacement, respectively.}
\label{fig.static_R}
\end{figure}

To quantitatively test our agreement between the discrete (vertex model) and the continuum description, we analyze three experimentally interesting and measurable observables for different active patch sizes $R$. i) maximum contraction ($\Delta A_{\text{max}}$), ii) maximum horizontal displacement ($u_{\text{max}}$), and iii) the position at which the displacement is maximum ($x_{\text{max}_d}$). In Fig.~\ref{fig.static_R}, we compare the observables in the final steady state, under medial $\left(\lambda_A=0.5\right)$ and perimeter $\left(\lambda_P=0.1\right)$ activity, finding an excellent agreement even for small values of $R$ ($R \sim $ cell size). We consider different intensities  for the activities, with $\lambda_A>\lambda_P$, because the ratio between the active forces goes  like $f_{A_0}/f_{P_0}\sim \lambda_A/ \left[\lambda_P (K_P/(K_A A_0)) (P_0/\sqrt{A_0})(c_2/\sqrt{c_1})\right]$, and since $K_P/\left(K_A A_0\right)=1$, $P_0/\sqrt{A_0}=3.9$, and $c_2/\sqrt{c_1}=2.63$ for the hexagonal case, $f_{A_0}/f_{P_0}\sim \lambda_A/ (10 \lambda_P)$. 
As a results of the areas and perimeter approximations, the continuum description slightly under predicts  the maximum contraction and horizontal displacement (black lines above mean values in Fig.~\ref{fig.static_R}A-B). Nevertheless, the results for the maximum contraction obtained from the continuum description lie inside the error bars of the discrete simulations.
Consistent with Fig.\  \ref{fig.R4_desordenado}, the change in area presents larger fluctuations for medial activity compared to activity in the perimeter. 
Interestingly, the standard deviations of the maximum horizontal displacement $u_{\text{max}}$ and the values of $x_{\text{max}_d}$, are remarkably similar for both activities. 
The cases of an active circular region (App.~\ref{app.circular}) and a tissue made of square cells (App.~\ref{app.squares}) present similar results.

\begin{figure}[t]
 \includegraphics[width=1\linewidth]{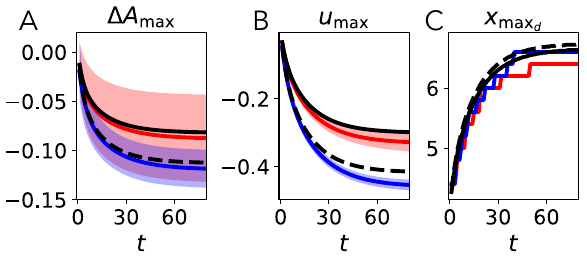} 
\caption{Comparison of micro and macro descriptions of the tissue evolution in time. (A) Maximum contraction in time. (B) Maximum horizontal displacement. (C) Position at which the horizontal displacement is maximum. Black curves show the outcomes of the continuum description, for medial $\left(\lambda_A=0.5,\lambda_P=0, R=4\right)$ (solid-line) and perimeter $\left(\lambda_A=0,\lambda_P=0.1, R=4\right)$ (dashed-line) activity, with $\tilde{\mu}=2\mu$. Solid-red and solid-blue curves represent the numerical mean results obtained from 100 simulations with a different active axis. Shaded areas represent standard deviations of $\Delta A$ and $u$ at the position of maximum contraction and displacement, respectively.}
\label{fig.evolution_R4}
\end{figure}

\subsection{Temporal evolution} \label{sec.temporal}
 
We numerically solve Eq.~\eqref{eq.temporal_stripe} imposing $u_x(x,t=0)=0$, $u_x(x=0,t)=0$, and  $u_x(x=W_{\text{half-box}},t)=0$. Since the original tissue obeys periodic boundary conditions, Euler's characteristic formula is given by $V-E+C=0$ ($V$: vertices, $E$: edges, $C$: cells). Additionally, in our tissue, vertices are formed by the intersection of three junctions, and then the coordination number is $2E/V=3$. By using both equations, we obtain the ratio $C:V = 1:2$, and then the area occupied by each vertex that appears in Eq. \eqref{eq.temporal_stripe} is $A_v=1/2$, in units of $A_0$.

We compare the temporal evolution of the observables between both descriptions in Fig.~\ref{fig.evolution_R4}, finding an excellent agreement. 
For both activities, $\sigma_{u_{\text{max}}}$ is small and very similar, showing independence with respect to the sign of the change (increase/decrease) of shape index $p_0$. Instead, $\sigma_{\Delta A_{\text{max}}}$ does depend on $p_0$, and is larger for medial activity, i.e., tissues with a core that has been actively fluidized (increased $p_0$). The numerical simulations using a disordered tissue allow us to obtain the dynamic evolution of each observable variance, paving the road for a theory beyond the mean field that depends on the tissue fluidity.
For the temporal evolution of the disordered tissue under a circular active region, and the tissue made of square cells under a simple active stripe, see App.~\ref{app.circular} and App.~\ref{app.squares}, respectively.

\begin{figure}[t]
 \includegraphics[width=1\linewidth]{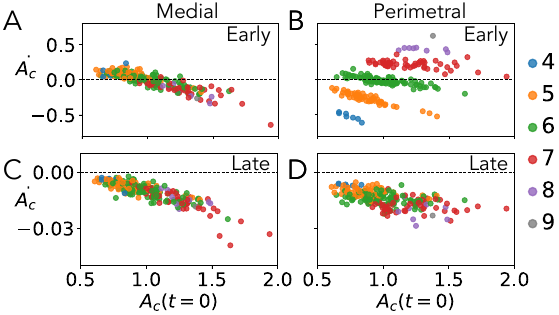} 
\caption{Cell-level growth rate. Left: simple active stripe with medial activity $\left(\lambda_A=0.5,\lambda_P=0, R=4\right)$. The same simulation is used in Fig.~\ref{fig.ejemplo_medial_R4}. Right: simple active stripe with perimeter activity $\left(\lambda_A=0,\lambda_P=0.1, R=4\right)$. We consider the 221 cells whose centers are initially at a distance smaller than $R/2$ from the active axis. The rate for each cell is defined as $\dot{A_c} = \left[A_c({t+\Delta})-A_c({t})\right]/\Delta$, with $\Delta=0.1$. (A)-(B) Scatter plot of the 221 rates versus the initial cell area $A_c(t=0)$, at $t=0.1$. (C)-(D) Scatter plot of the 221 rates versus the initial cell area $A_c(t=0)$, at $t=1$. The colors indicate the number of sides of each cell.}
\label{fig.A_rates}
\end{figure}

The non-uniform distributions of $A_{0c}$ and $P_{0c}$ in the numerical simulations add an extra source of mechanical heterogeneity. We measure the rate of change of the area of cells whose centers are initially at less than $R/2$ from the active axis. For these cells $\Gamma(x)\in [0.88,1]$ is rather homogeneous, and then heterogeneity in activity comes mainly from the differences in $A_{0c},P_{0c}$.
The cell growth rates show two different regimes, namely for early times ($t=0.1$) and late times ($t=1$).  Fig.~\ref{fig.A_rates} shows these area rates for the case for medial activity ($\lambda_A=0.5, \lambda_P=0, R=4$, same simulation shown in Fig.~\ref{fig.ejemplo_medial_R4}), and perimeter activity ($\lambda_A=0, \lambda_P=0.1, R=4$). At early times, for both activities, we find that some cells grow, while other shrink. For late times, all the cells are shrinking. 
Additionally, the distribution of growing/shrinking cells at short times depends on the kind of activity. For medial activity, big cells, which happen to have high sidedness ($\geq 6$), contract first, with a rate that increases with the cell size, Fig.~\ref{fig.A_rates}A. For perimeter activity, cells with low sidedness ($\leq 6$) contract first. However, inside each family of polygons the contraction rate increases with size as in the medial activity case, creating the clustering in lines observed in Fig.~\ref{fig.A_rates}B.
The continuum description of the tissue is consistent with these results. Indeed, in the continuum formulation, the contraction rate is proportional to the rate of deformation, which has a global size dependence (as seen in Eq.~\eqref{eq.temporal_stripe}), as well as specific dependences originating from the different forces. Specifically, the active forces (second and third terms in Eq.~\eqref{eq.eq_x}) have the pre-factors $A_{0}$ for the medial case and $P_{0}(c_2/c_1)$ for the perimeter case. We find that in our isotropic disordered tissue $A_{0c}$ and $P_{0c}$ increase with cell size in the initial configuration (Fig.~\ref{fig.initial_state}C-D), whereas the term $(c_2/c_1)$ decreases with the number of sides of a given polygon Fig.~\ref{fig.initial_state}B.
Since our tissue is initially isotropic ($c(t=0)\approx 0$ for all cells), we see from Eqs.~\eqref{eq.p_P0} and \eqref{eq.dev_P0} that perimeter active pressure initially dominates over the perimeter active tensions, causing cellular contraction and shape deformations, changing $c$ and hence increasing active tensions. Since $\sqrt{\cosh c}$ is an even function and $\sinh c$ is an odd function, $\sigma_{{P_0},\text{dev}}$ increases if elongated ($c>0$), and decreases if contracted  ($c<0$). Then, the active tension regulates forces in order to increase isotropy.

\section{Conclusions} \label{sec.discussion}
We developed a continuum model for confluent epithelial tissues under spatial inhomogeneous cellular activity, constructed from the stress tensor of the cell-level vertex model. Here, activity directly changes the target cell shape index $p_0$, which has been shown to control fluidity in homogeneous tissues \cite{bi2015density}. 
Our continuum model shows that active changes in the target cell shape index are directly linked to cellular stresses. Activity in the medial region of the cell generates an isotropic pressure (explicitly) independent of shape, while activity in the perimeter region of the cell generates an isotropic pressure and a deviatoric (trace-less) stress, both of them with an explicit dependence on shape and size. Particularly, the active deviatoric stress acts against the shape anisotropy and vanishes under isotropic configurations. 

To test the validity of the continuum description, we compared the steady-state and time-dependent solutions with computer simulations of isotropic disordered tissues using the vertex model. We used inhomogeneous target areas and perimeters to represent inhomogeneous patches presenting apical contractions. Particularly, we analyzed the case of a simple active stripe, where activity decreases with distance from an active axis, proportional to a Gaussian function. 
Remarkably, our continuum formulation that uses only two degrees of freedom to define shape, presents excellent agreement when using the exact cell-level parameters of the vertex model simulations.
Similar results are reported when considering a circular active region (App.~\ref{app.circular}) and a simple active stripe over a tissue made of square cells (App.~\ref{app.squares}). 

T1 topological events, which drive relative cellular motion, were unnecessary here since the levels of activity used in this work generate small deformations. Then, all tissues considered in this work are in a solid state by definition. Nevertheless, numerical simulations show that different levels of geometrical variation arise as a non-dynamic quantification of how cellular activity affects tissue fluidity. As a liquid versus solid behavior, medial activity (increasing $p_0$) produces more significant cell-level geometrical variations than perimeter activity (decreasing $p_0$). Interestingly, no significant difference is seen at the vertex level. 

Additionally, we found that heterogeneous preferred areas and perimeters in the vertex model induce mechanical heterogeneity even when the active functions are mostly uniform. In particular, we measured the rate of cell contraction/expansion at short times for cells close to the active axis. Since our tissues are initially isotropic, active pressures dominate. The active force under medial activity depends on the size only, and particularly bigger cells contract first. Instead, under perimeter activity, the active force inducing contraction depends on both polygon size and sidedness. We find that cells with lower sidedness contract first, while maintaining the positive correlation between cell size and rate of contraction.
In the second case, cellular contraction in a confluent system leads to shape changes, turning on the active tensions that increase isotropy in the system. The latter is related to experimental observations and computational models, showing that more compact cells tend to be more solid (lower $p_0$), whereas elongated cells increase the tissue fluidity (larger $p_0$) \cite{park2015unjamming,wang2020anisotropy,damavandi2022universal,bi2015density}.

Further research directions include examining the fluid/solid dynamic disparities caused by inhomogeneous cellular activity, spatial and temporal, in motile tissue models and analyzing the cell level shape dynamic in anisotropic pre-stressed tissues. Also, rules for cellular division and extrusion could be considered, as in Ref.~\cite{kursawe2015capabilities}, where they can explain different extrusion rates caused by similar activities. 

\appendix

\section{Elastic stress tensor derivation} \label{app.stress}

To obtain an elastic stress that depends on geometrical shapes, we use the texture matrix [Eq.~\eqref{eq.M}]. Then, the area and perimeter of the cell can be written as $A_c= c_1 \sqrt{\det{\mathbb{M}_c}}$ and $P_c=c_2\sqrt{\tr{\mathbb{M}_c}}$, respectively, with $c_1$ and $c_2$ depending on the type of polygon. Hence, the energy functional [Eq.~\eqref{eq.Evertex}] for a tissue with periodic boundary conditions turns to be equal to
\begin{align}
E = & \frac{K_A}{2}\sum_c\left( c_1 \sqrt{\det{\mathbb{M}_c}} - A_{0c}\right)^2 + \frac{K_P}{2}\sum_c\left[ c_2\sqrt{\tr{\mathbb{M}_c}} - \left(P_{0c}-\frac{J}{2K_P}\right)\right]^2.
\label{eq.Evertex2}
\end{align}

We perform a coarse-graining over the discrete perspective of the tissue to obtain a smooth symmetric tensor field $\mathbb{M}\left(\VEC r\right)$, which represents the shape at a tissue scale. We define the energy density per unit area [Eq.~\eqref{eq.f.maintext}], such that the total elastic energy is $E=\int f\left(\mathbb{M}\left(\VEC r\right)\right){\rm d}^2 r$, as in Ref.~\cite{ishihara2017cells}.  Analogous to Ref.~\cite{ishihara2017cells}, we compute the elastic stress as $\sigma_e = f\mathds{1} + 2\left(\partial f/ \partial \mathbb{M}\right)\mathbb{M}$. Using that $\partial \tr{\mathbb{M}}/ \partial \mathbb{M}=\mathds{1}$ and $\partial \det{\mathbb{M}}/ \partial \mathbb{M}=(\det{\mathbb{M}}) \mathbb{M}^{-1}$, the elastic stress tensor is finally given by 
 \begin{align}
\sigma_e =&K_A\left(c_1\sqrt{\det{\mathbb{M}}}-A_0\right)\mathds{1}\nonumber \\
&+ \frac{c_2 K_P}{c_1\sqrt{\det{\mathbb{M}}}\sqrt{\tr{\mathbb{M}}}}\left[c_2\sqrt{\tr{\mathbb{M}}}-\left(P_{0}-\frac{J}{2K_P}\right)\right]\mathbb{M},\nonumber \\
=& K_A\left(c_1M-A_0\right)\mathds{1}+K_P\frac{c_2}{c_1}\left(c_2 -\frac{P_0 - J/(2K_P)}{\sqrt{2M\cosh(c)}}\right)\frac{\mathbb{M}}{M},
 \label{eq.sigmae}
\end{align}
where we used that for $\mathbb{M} = M e^{c\Theta}$,  $\det{\mathbb{M}}=M^2$ and $\tr{\mathbb{M}} = 2M\cosh\left(c\right)$.
Since $\mathbb{M} = M e^{c\Theta}=M\cosh(c)\mathds{1} + M\sinh(c)\Theta$, we can rewrite the last term in Eq.~\eqref{eq.sigmae}, identifying the pressure [Eq.~\eqref{eq.p_e1}] and a deviatoric (trace-less) elastic stress [Eq.~\eqref{eq.dev_e1}].

\section{Simulations details} \label{app.sim.details}

\begin{figure}[t!] 
\centering 
 \includegraphics[width=0.9\linewidth]{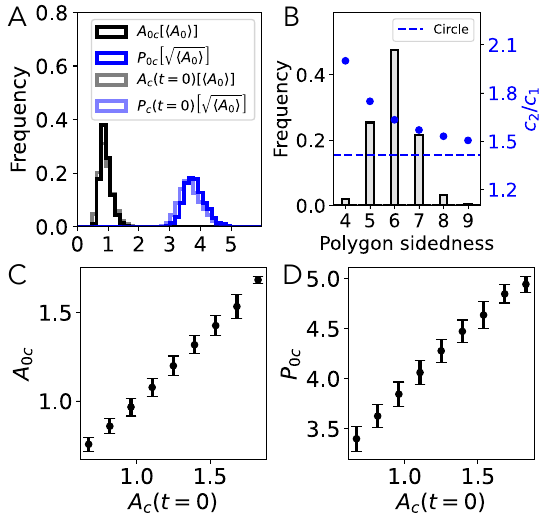} 
\caption{Initial configuration of the relaxed disordered tissue. (A) Histograms of the cell target area, target perimeter, initial area, and initial perimeter. $\langle A_{0c}\rangle=\langle A_{c}\rangle=1, \langle P_{0c}\rangle=3.9, \langle P_{c}\rangle=3.77$. (B) Histogram of polygon sidedness in gray bars. Ratio $c_2/c_1$ as a function of polygon sidedness, for regular polygons (blue dots). The dashed line represents the value for a circle (limit of an infinite number of sides). (C) Correlation between target area and initial area. (D) Correlation between target perimeter and initial area. Error bars represent $\pm 1$ standard deviation.
}
\label{fig.initial_state}
\end{figure}

The initial (relaxed) configuration of our entire disordered tissue in the numerical simulations, has polygonal cellular shapes from squares to nonagons, with majority of hexagons, and  $\langle A_{c}\rangle=1$ and $ \langle P_{c}\rangle=3.77$ (Fig.~\ref{fig.initial_state}).

For the case of the active stripe in the disordered tissue, for each simulation we randomly choose a cell whose horizontal position defines the position of the active axis. Then, through the simulations we keep fixed all the vertices that belong to cells whose centers lie further away than $W_{\text{half-box}}=20$, from the active axis. In each simulation with an active stripe, the effective tissue has $\sim 2200$ cells, vertical periodic boundary conditions, and horizontal fixed boundary conditions.

For the case of the circular active patch in the disordered tissue, for each simulation we randomly choose a cell whose center defines the origin of activity. Then, through the simulations we keep fixed all the vertices that belong to cells whose centers lie further away than $W_{\text{half-box}}=24$ from the active origin. In each simulation with a circular active patch, the effective tissue has $\sim 1800$ cells, and fixed boundary conditions.

Every simulation is run using the parameters from Table~\ref{tab.parameters_stripe}, with $W_{\text{half-box}}=\lbrace{20,24,14\rbrace}$ for the disordered and regular tissue with a simple active stripe, for the disordered and regular tissues with a circular active patch, and for the tissue made of square cells with a simple active stripe, respectively.

\begin{table}[h]
\small
  \caption{\ Simulations parameters}
  \label{tab.parameters_stripe}
  \begin{tabular*}{0.48\textwidth}{@{\extracolsep{\fill}}lll}
    \hline
    Parameter & Symbol & Value \\
    \hline
    Area elastic modulus & $K_A$ & 1  \\ 
    Perimeter elastic modulus & $K_P$ & 1  \\ 
    Constant line tension & $J$ & 1  \\ 
    Mean target area & $A_0=\langle A_{0c}\rangle $ & 1  \\ 
    Friction coefficient & $\mu$ & 1 \\ 
    Simulation box width & $L_x$ & $53.73$ \\ 
    Simulation box length & $L_y$ & $ 55.84$  \\ 
    Active region size & $R$ & $1$--$20$  \\ 
    Box half-width active simulation & $W_{\text{half-box}}$ & $\lbrace{20,24,14\rbrace}$  \\ 
    Medial activity parameter & $\lambda_A$ & $\lbrace{0,0.5\rbrace}$  \\ 
    Perimeter activity parameter & $\lambda_P$ & $\lbrace{0,0.1\rbrace}$  \\ 
    Integration time step & $\Delta t$ & $0.05$ \\
    \hline
  \end{tabular*}
\end{table}

 \section{Circular active patch}\label{app.circular}

\subsection{Ordinary differential equation system}
We consider a tissue initially at equilibrium characterized by a uniform field $\mathbb{M}$, with $M(t=0) = m = A_0/c_1$, $c(t=0) = 0$ and $\theta(t=0)=0$, which is subject to an active process that changes radially the target parameters as $A_{0} \rightarrow  A_{0} + \Delta_A\left(\VEC r\right)$ and $P_{0} \rightarrow P_{0}+ \Delta _P \left(\VEC r\right)$, with $\Delta_A= -\lambda_A A_{0c}\Gamma(r) $ and $\Delta_P= -\lambda_P P_{0c}\Gamma(r)$. Here, $\Gamma(r)= \exp\left(-r^2/R^2\right)$ is the function that describes the active inhomogeneity in the tissue, where $r$ is the radial coordinate measured from the center of the circular patch, and $R$ represents the active region extension. Positive values for $\lambda_A$ and $\lambda_P$  cause apical contractions. In the case of a circular patch, the active stresses are given by  $\sigma_{A_0} = -p_{A_0}\mathds{1}$ and $\sigma_{P_0}  = -p_{P_0}\mathds{1} + \sigma_{P_0,\text{dev}}$, with pressures and the deviatoric stresses as in Eqs.~\eqref{eq.p_A0}, \eqref{eq.p_P0}, and \eqref{eq.dev_P0}.
Given the symmetries of our initial system, we assume a radial displacement field $\VEC u = u_r(r)\HAT r$, with the coarse-grained orientations given by the $\theta$-polar angle, and the scalar fields depending only on $r$, $M = M(r)$ and $c = c(r)$. Then, the confluence condition for the tissue [Eq.~\eqref{eq.confluence}], in polar coordinates, reads

\begin{figure}[!b] 
\centering 
 \includegraphics[width=.9\linewidth]{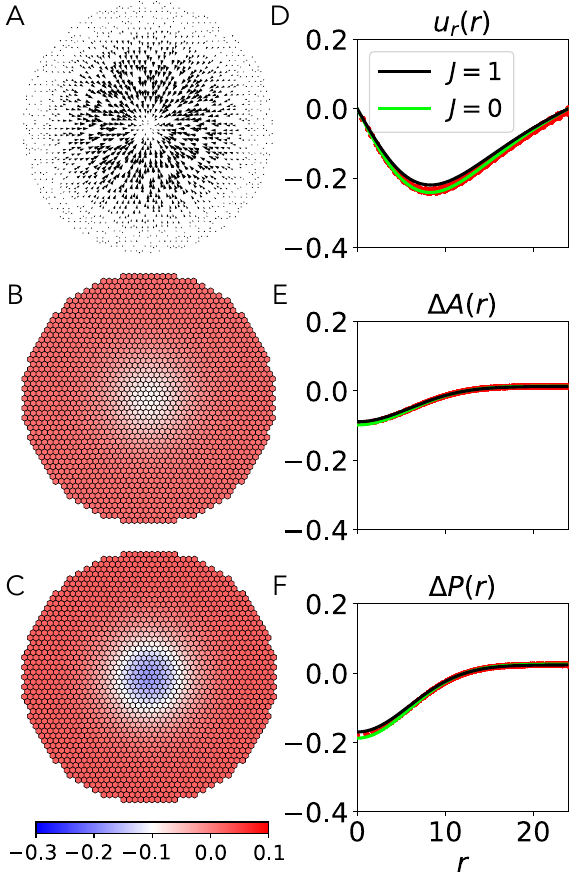} 
\caption{Contraction of a circular active patch in an ordered tissue. Results of the simulated circular  epithelial tissue, made from regular hexagonal cells, with a circular active patch under medial activity ($\lambda_A=0.5$, $\lambda_P=0$, $R=6$, $P_0=3.72$, and $W_{\text{half-box}}=24$) at $t=100$, versus the initial position of the vertex or cell centers. (A) Vectorial map of total vertex displacement in the complete effective tissue, in units of $\left[\sqrt{A_0}/5\right]$. (B) Area change for the whole, effective tissue. (C) Perimeter change for the whole, effective tissue. (D) Scatter plot of the radial displacement of vertices (red dots). (E) Scatter plot of the area change (red dots). (F) Scatter plot of the perimeter change (red dots). The solid-black (lime) curves show the outcomes of the continuum description, using $J=1$ ($J=0$), $c_1=3\sqrt{3}/2$, $c_2=3\sqrt{2}$, $A_0=1$, and $P_0=3.72$.
}
\label{fig.circular-hexagons}
\end{figure}

 \begin{align}
M\begin{pmatrix}
e^c & 0 \\
0 & e^{-c} 
\end{pmatrix} = m
\begin{pmatrix}
1+2\frac{{\rm d} u_r}{{\rm d}  r} & 0 \\
0 & 1+ 2\frac{u_r}{ r}
\end{pmatrix},
 \label{eq.confluent}
 \end{align}
 from where we obtain the following relations
  \begin{align}
  &M\left(r\right)=m\sqrt{\left(1+2\frac{{\rm d} u_r}{{\rm d}  r}\right)\left(1+ 2\frac{u_r}{ r}\right)}, \label{eq.M0_u}\\
&\cosh\left[c\left(r\right)\right] = \frac{m}{M\left(r\right)}\left( \frac{{\rm d} u_r}{{\rm d}  r} +\frac{u_r}{ r}+1 \right)
=\frac{\frac{{\rm d} u_r}{{\rm d}  r} +\frac{u_r}{ r}+1}{\sqrt{\left(1+2\frac{{\rm d} u_r}{{\rm d}  r}\right)\left(1+ 2\frac{u_r}{ r}\right)}},\label{eq.c1_u}\\
&\sinh\left[c\left(r\right)\right] = \frac{m}{M\left(r\right)}\left( \frac{{\rm d} u_r}{{\rm d}  r} -\frac{u_r}{ r}\right)
=\frac{\frac{{\rm d} u_r}{{\rm d}  r} -\frac{u_r}{ r}}{\sqrt{\left(1+2\frac{{\rm d} u_r}{{\rm d}  r}\right)\left(1+ 2\frac{u_r}{ r}\right)}}.
 \label{eq.c2_u}
 \end{align}

Since all the fields depend on the radial coordinate only, each stress term ($\sigma_i=\lbrace{ \sigma_e , \sigma_{\text{$A_0$}}, \sigma_{\text{$P_0$}} \rbrace}$) generates a radial force
  \begin{align}
 \nabla \cdot \sigma_i=& \left( \frac{\partial \sigma_i^{rr}}{\partial r} + \frac{1}{r}\left[ \sigma_i^{rr} -  \sigma_i^{\theta \theta} \right]\right) \HAT r = f_i \HAT r ,
 \label{eq.fi} 
\end{align}
 where 
\begin{align}
\sigma_e^{rr}=&K_A\left(c_1M-A_0\right)+K_P\frac{c_2}{c_1}\left(c_2 -\frac{P_0 - J/(2K_P)}{\sqrt{2M\cosh[c(x)]}}\right)\sqrt{\frac{1+2\frac{{\rm d} u_r}{{\rm d}  r}}{1+2\frac{ u_r}{  r}}},\\
\sigma_e^{\theta\theta}=&K_A\left(c_1M-A_0\right) + K_P\frac{c_2}{c_1}\left(c_2 -\frac{P_0 - J/(2K_P)}{\sqrt{2M\cosh[c(x)]}}\right)\sqrt{\frac{1+2\frac{ u_r}{  r}}{1+2\frac{{\rm d} u_r}{{\rm d}  r}}}.
\end{align}
Then, the equilibrium condition is written as $\sum_i f_i = 0.$ Finally, we can write each force [Eq.~\eqref{eq.fi}] in terms of $r$,  $u_r\left(r\right)$,  $u_r'\left(r\right)$, and  $u_r''\left(r\right)$, obtaining an ordinary differential equation from the equilibrium condition. Analogous to the simple active stripe case,  we numerically solve the equation considering the previously described initial condition, imposing $u_r(r=0)=0$, $u_r(r=W_{\text{half-box}})=0$, and using $c_1=3\sqrt{3}/2$, $c_2=3\sqrt{2}$. To analyze the temporal evolution, we numerically solve 
  \begin{align}
\frac{\mu}{A_v}\frac{{\partial}u_r(r,t)}{{\partial }t} = \sum_i f_i(r,t),
 \label{eq.temporal_stripe2}
 \end{align}
 where $f_i=\lbrace{f_e , f_{\text{$A_0$}}, f_{\text{$P_0$}} \rbrace}$, $A_v=1/2$, imposing $u_r(r,t=0)=0$, $u_r(r=0,t)=0$, and  $u(r=W_{\text{half-box}},t)=0$.

\subsection{Circular active patch: Results}
\begin{figure}[t] 
\centering 
\includegraphics[width=.9\linewidth]{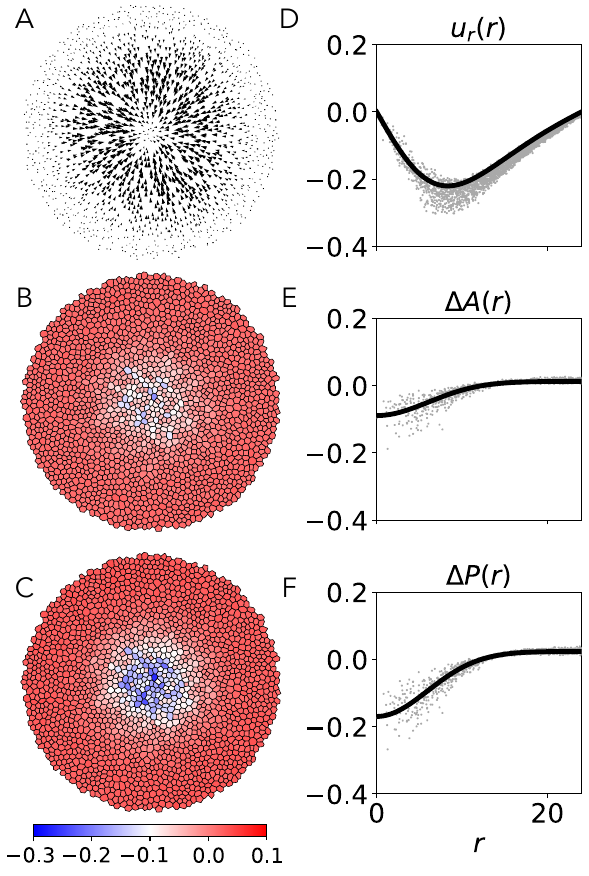} 
\caption{Contraction of a circular active patch in a disordered tissue. Results of the simulated circular disordered epithelial tissue with a circular active patch under medial activity ($\lambda_A=0.5$, $\lambda_P=0$, $R=6$, and $W_{\text{half-box}}=24$) at $t=100$, versus the initial position of the vertex or cell centers. (A) Vectorial map of total vertex displacement in the complete effective tissue, in units of $\left[\sqrt{A_0}/5\right]$. (B) Area change for the whole, effective tissue. (C) Perimeter change for the whole, effective tissue. (D) Scatter plot of the radial displacement of vertices. (E) Scatter plot of the area change. (F) Scatter plot of the perimeter change. The solid-black curves show the outcomes of the continuum description, using  $c_1=3\sqrt{3}/2$, $c_2=3\sqrt{2}$, $A_0=1$, and $P_0=3.9$.
}
\label{fig.circular-disordered}
\end{figure}

Figure \ref{fig.circular-hexagons} shows the vertex displacement, change in the cellular area, and change in cellular perimeter, for an ordered tissue (regular hexagonal cells) of radius $W_{\text{half-box}}=24$ under fixed boundary condition, with a circular region ($R=6$) under medial activity defined by $\lambda_A=0.5,\lambda_P=0$. 
As in Fig.~\ref{fig.hex-franja}, we find that the results from continuum description depend on $J$ (solid-black vs lime curve in Fig.~\ref{fig.circular-hexagons}D-E-F), in contrast to the numerical simulations for this high-symmetry case. 
Fig.~\ref{fig.circular-disordered} shows the analogous results when considering the disordered tissue. Vertices displacements are approximately in the radial direction, and geometrical changes in area and perimeter show large variations near the origin of the activity. Therefore, the maximum contraction is not well defined and is not considered as a relevant experimental observable to quantify. The results obtained from the continuum description, with $J=0$, lie inside the gray data points obtained from numerical simulations using the two-dimensional vertex model.

\begin{figure}[t] 
\centering 
 \includegraphics[width=0.8\linewidth]{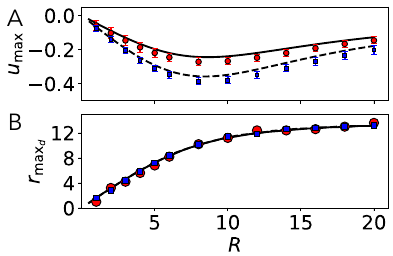} 
\caption{Comparison of micro and macro descriptions of the steady-state. Results of the simulated circular disordered epithelial tissue with a circular active patch under circular activity ($R=6,W_{\text{half-box}}=24$). (A) Maximum horizontal displacement. (B) Radius at which the radial displacement is maximum. Black curves show the outcomes of the continuum description for medial $\left(\lambda_A=0.5,\lambda_P=0.0\right) $(solid-line) and perimeter $\left(\lambda_A=0,\lambda_P=0.1\right)$ (dashed-line) activity. Red circles and blue squares represent the numerical results averaged over 100 simulations with different active origins for different integer $R$ between 1 and 20. Bars show the standard deviations of $u$ at the position of maximum displacement.
}
\label{fig.S3}
\end{figure}

\begin{figure}[h] 
\centering 
 \includegraphics[width=0.8\linewidth]{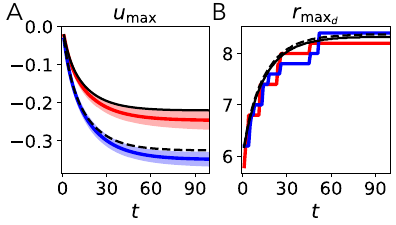} 
\caption{Comparison of micro and macro descriptions of the tissue evolution in time. (A) Maximum horizontal displacement. (B) Radius at which the radial displacement is maximum. Black curves show the outcomes of the continuum description with $\tilde{\mu}=2\mu$ (hexagonal lattice), for medial ($\lambda_A=0.5$, $\lambda_P=0$, and  $R=6$) (solid-line) and perimeter ($\lambda_A=0$, $\lambda_P=0.1$, and  $R=6$) (dashed-line) activity. Solid-red and solid-blue curves represent the numerical mean results obtained from 100 simulations with different active axes. Shaded areas represent standard deviations of $u_r$ at the radius of maximum displacement.}
\label{fig.S4}
\end{figure}

Figure \ref{fig.S3} shows the maximum vertex radial displacement and the radius at which it occurs (measured with respect to the origin of activity) in the final steady-state, considering medial activity with ($\lambda_A=0.5$, $\lambda_P=0$), and perimeter activity with ($\lambda_A=0$, $\lambda_P=0.1$) using the disordered tissue. The results obtained from the continuum description ($c_1=3\sqrt{3}/2$,  and $c_2=3\sqrt{2}$) lie inside the error bars of the numerical simulations. Analog to the simple active stripe, the standard variation of the radial displacement at the position of maximum displacement is small and similar for both activities. Also, the position of maximum displacement is remarkably similar for both kinds of activity. Figure \ref{fig.S4} shows the temporal evolution of the maximum vertex radial displacement, the radius at which it occurs, and its agreement with the continuum description (using $c_1=3\sqrt{3}/2$, $c_2=3\sqrt{2}$, and $\tilde{\mu}=2\mu$).

\section{Tissue made of square cells with a simple active stripe}\label{app.squares}

\begin{figure}[h] 
\centering 
 \includegraphics[width=1\linewidth]{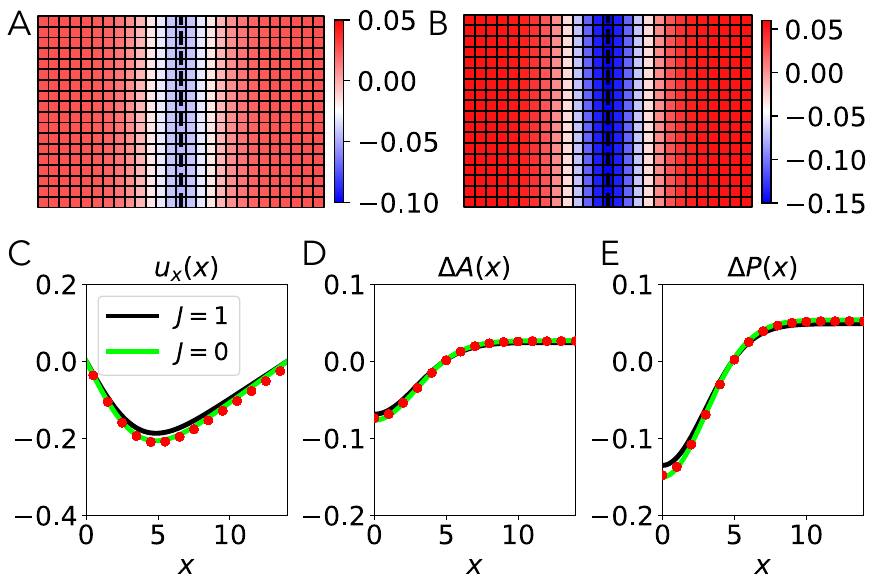} 
\caption{Contraction of a simple active stripe: square cells. Results of the simulated epithelial tissue made of initially square cells of side equals one, with $K_A=1$, $K_P=1$, $A_0=1$ , $P_0=4$, and $J=1$, under medial activity ($\lambda_A=0.5$, $\lambda_P=0$,  $R=3$, and $W_{\text{half-box}}=14$) at $t=100$, versus the initial position of the vertex or cell centers. (A) Representative section of the tissue, showing area change. (B) Representative section of the tissue, showing perimeter change. (C) Scatter plot of the horizontal displacement of vertices (red dots). (E) Scatter plot of the area change (red dots). (F) Scatter plot of the perimeter change (red dots). The solid-black (lime) curves show the outcomes of the continuum description, with $J=1$ ($J=0$), $A_0=1$, $P_0=4$, $c_1=2$, and $c_2=4$ (square cells).
}
\label{fig.S5}
\end{figure}

 As an additional test of the validity of the continuum description, we consider a tissue made of regular square cells, for which $c_1$, $c_2$, and $\tilde \mu$ change in comparison with the case of hexagonal cells.  Fig.~\ref{fig.S5} shows the vertex displacement, change in the cellular area, and change in the cellular perimeter for a tissue of size $L_y$ and $W_{\text{half-box}}=14$, made of square cells only. The tissue is under horizontal fixed boundary conditions and vertical periodic boundary conditions. We consider an active stripe defined by $R=3$ under medial activity with $\lambda_A=0.5,\lambda_P=0$, with $K_A=1, K_P=1$, and $J=1$. Due to the high symmetry, vertices are horizontally displaced only, and there is no standard deviation for each quantity. As for the simulations shown in Figs.~\ref{fig.hex-franja} and \ref{fig.circular-hexagons}, here the value of $J$ is irrelevant in the vertex model simulation due to the high symmetry. Fig.~\ref{fig.S5} shows the $J$-dependence of the continuum description. Overall, the agreement is excellent.

\begin{figure}[h] 
\centering 
 \includegraphics[width=0.95\linewidth]{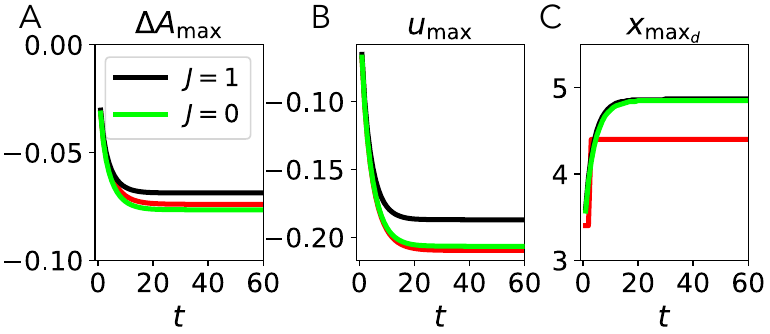} 
\caption{Comparison of micro and macro descriptions of the tissue evolution in time. Results of the simulated epithelial tissue made of initially regular square cells of side one, with $K_A=1, K_P=1$, $A_0=1$, $P_0=4$, and $J=1$, under medial activity ($\lambda_A=0.5,\lambda_P=0, R=3,W_{\text{half-box}}=14$) at $t=100$. (A) Maximum contraction in time. (B) Maximum displacement in time. (C) Position at which the displacement is maximum in time. The solid-black (lime) curves show the outcomes of the continuum description, with $J=1$ ($J=0$), $A_0=1$, $P_0=4$, $c_1=2, c_2=4, \tilde{\mu}=\mu$ (square lattice).
}
\label{fig.S6}
\end{figure}

 Figure \ref{fig.S6} shows the temporal evolution of the maximum contraction (which takes place at $x=0$), the maximum vertex displacement, and the position at which it occurs, along with the continuum solution when using two values of $J$. Since cells are squares, we use $c_1=2, c_2=4$ in the continuum description. Also, since a square lattice has a ratio of cells over vertices of one, then  $\tilde{\mu}=\mu$. \\

\section*{Author Contributions}
Conceptualization, F.P.-V. and R.S.; methodology, F.P.-V. and R.S.; software, F.P.-V.; validation, F.P.-V. and R.S.; formal analysis, F.P.-V. and R.S.; investigation, F.P.-V.; resources, R.S.; writing-original draft, F.P.-V. and R.S.; writing-review and editing, F.P.-V. and R.S.; supervision, R.S.; funding acquisition, R.S.

\section*{Conflicts of interest}
There are no conflicts to declare.

\section*{Acknowledgements}
We thank  Shiladitya Banerjee for useful suggestions.
This research was supported by the Fondecyt Grant No.~1220536 and Millennium Science Initiative Program NCN19\_170D of ANID, Chile.


\balance



\providecommand*{\mcitethebibliography}{\thebibliography}
\csname @ifundefined\endcsname{endmcitethebibliography}
{\let\endmcitethebibliography\endthebibliography}{}

\end{document}